\documentclass[letter]{aa}
\usepackage{graphicx}
\usepackage{txfonts}
\usepackage{natbib}
\begin{document}

\title{On the distance of \object{PG 1553+11}}
\subtitle{A lineless BL Lac object active in the TeV band}

\author{ Aldo  Treves\thanks{Associated to INAF and INFN} \inst{1}\and Renato Falomo \inst{2} \and Michela Uslenghi \inst{3}
 }

\institute{University of Insubria, Via Valleggio 11, 22100 Como, Italy\\
\email{aldo.treves@uninsubria.it} \and INAF -- Osservatorio
Astronomico di Padova, Vicolo dell'Osservatorio 5, 35122 Padova,
Italy\\ \email{renato.falomo@oapd.inaf.it}
 \and
INAF -- IASF, Via Bassini 15, I-20133 Milano, Italy\\
\email{uslenghi@iasf-milano.inaf.it} }

\offprints{Michela Uslenghi}

\date{Received ..., ; accepted ...}

  \abstract
    {The redshift of \object{PG 1553+11}, a bright BL Lac object ($V\sim14$), is still unknown. It has been recently observed in the
       TeV band, a fact that offers an upper limit for the redshift $z<0.4$.
    }
   { We intend to provide a lower limit for the distance of the object.}
   {We used a $\chi^2$  procedure to constrain the apparent magnitude
    of the host galaxy in archived HST images. Supposing that the host galaxy is typical of BL Lac objects ($M_{R}$~-22.8),
    a lower limit to the distance can be obtained from the limit on
    the apparent magnitude of the host galaxy. }
   {Using the $3\sigma$ limit on the host galaxy magnitude, the
   redshift is found to be $\geq$ 0.25.}
   {The redshift of \object{PG 1553+11} is probably in the range z=0.3--0.4,
   making
   this object the most distant extragalactic source so far detected
   in the
   TeV energies. We suggest that
      other bright BL Lac objects of unknown redshift and similar
      spectroscopic characteristics
      may be interesting targets for TeV observations.}

\titlerunning{The distance of PG 1553+11 }
\authorrunning{Treves et al. }

   \maketitle
%

\section{Introduction}

Although having been studied with advanced instruments, the very
bright ($V\sim14$) BL Lac object \object{PG1553+11} still has no
line detected \citep[e.g.][and an example in Fig. 1]{03,12}. Based
on the hypothesis that the active nucleus sits in a typical host
galaxy \citep[a giant elliptical of $M_{R}\sim-23$, see sect. 2.2
and][]{11}, \cite{12} propose a lower limit to the redshift $z>0.1$
derived from the upper limit of the equivalent widths of absorption
lines.

\object{PG 1553+11} has been recently detected in the TeV band using
the atmospheric Cherenkov technique, both by the HESS and the MAGIC
collaborations \citep{00,01}. Because of the opacity due to
photon-photon interaction on the extragalactic background light
\citep[EBL, for a recent contribution on the subject and references,
see][]{08}, the very detection of the source in the TeV band implies
an upper limit to its distance. Its actual value depends on the
hypothesis on the intrinsic TeV spectral shape, and limits of
$z<0.8$ and $z< 0.4$ have been proposed \citep{07} that assume a
minimal EBL contribution. The most restrictive value corresponds to
the reasonable assumption of an intrinsic spectrum without
flattening and high--energy breaks.

\begin{figure}
\centering \resizebox{9cm}{6cm}{\includegraphics{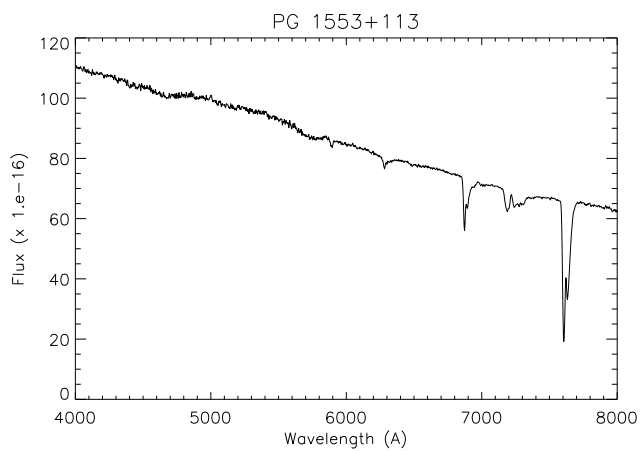}}
\caption{The optical spectrum of the BL Lac object \object{PG
1553+11} obtained with VLT and FORS. Apart from the telluric
absorptions, the spectrum consists of a non thermal featureless
emission \citep[see for details ][]{12}. Additional faint
absorptions at short wavelengths are due to molecular gas in our
galaxy. \label{Spectrum}}
\end{figure}

In this letter  we reanalyze the images obtained with the WFPC2
camera onboard HST in order to constrain the lower limit to the
redshift  (Sect. 2). Our results are then compared with those from
the TeV observations, and the overall astrophysical picture is
briefly discussed.


\section{The image of \object{PG 1553+11}}

\subsection{Analysis of the HST image }

The object was imaged by the PC camera (0.046 arcsec/pixel) of
HST/WFPC2, observed with the F702 filter for 610 sec, as part of a
program aimed at systematically studying the host galaxies of BL Lac
objects \citep{13,14}. The HST image, obtained by combining 3 images
with different time exposures, is reported in Fig. 2. \citet{13} and
\citet{14} did not find any evidence of a host galaxy, confirming
previous indications from ground--based observations \citep{04}. The
object was noticed because the absence of the host galaxy was
combined with a high nuclear brightness. We analyzed the combined
image with AIDA (Astronomical Image Decomposition and Analysis), a
software package for 2-d model fitting, specifically designed to
measure host galaxies of AGN \citep{15}.
\begin{figure}
\centering \resizebox{8cm}{8cm}{\includegraphics{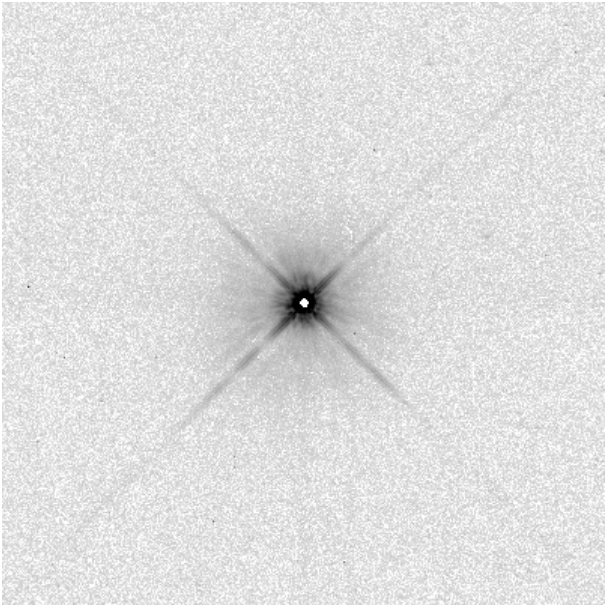}}
\caption{ The optical (R filter) image of the BL Lac object PG
1553+11 obtained by HST + WFPC2 (a 18.3$\times$18.3
 $arcsec^{2}$ box is shown). \label{Image}}
\end{figure}

The BL Lac image was prepared for the analysis by building a mask to
exclude any residual bad pixels. Because the image of the target is
saturated, the central core of the object (up to 0.2 arcsec) was
also masked. The local background was estimated from the signal's
average level computed in a circular annulus centered on the object
(8-9 arcsec).

The PSF was modeled using the PSF generated by Tiny Tim \citep{06}
at the object's location in the core (within 1.5 arcsec). However,
it is known that Tiny Tim does not properly model the external faint
halo produced by the scattered light. To take this effect into
account, an exponential component was added to the external part,
with a smooth transition (1.5-3.5 arcsec) between the pure Tiny Tim
and the mixed PSF. An archive image of a star produced with the same
instrumental setup was used to constraint model parameters by
fitting.

The BL Lac image was fitted with the PSF model, showing that the
radial profile of the object agrees well with the PSF, and no
deviation is seen at large radii. Thus, the object is not resolved
(in agreement with previous analysis carried out on this image). In
Fig. 3 we report the azimuthally--averaged radial profile and
compare it with the profile of a scaled PSF. The overall agreement
is very good, but some small deviations at radii smaller than
$\sim$1.5 arcsec are apparent in the plot of the residuals (see
lower panel of Fig.3). These wave-shaped deviations from the pure
TinyTim model are very likely due to the undersampling of the HST
PSF and to the ``breathing'' of HST that produces a slight change in
the PSF shape at the focal plane. This shows the importance of
systematic effects and casts some doubt on the suitability of the
$\chi^{2}$ statistics.

\begin{figure}
\centering \resizebox{9cm}{11cm}{\includegraphics{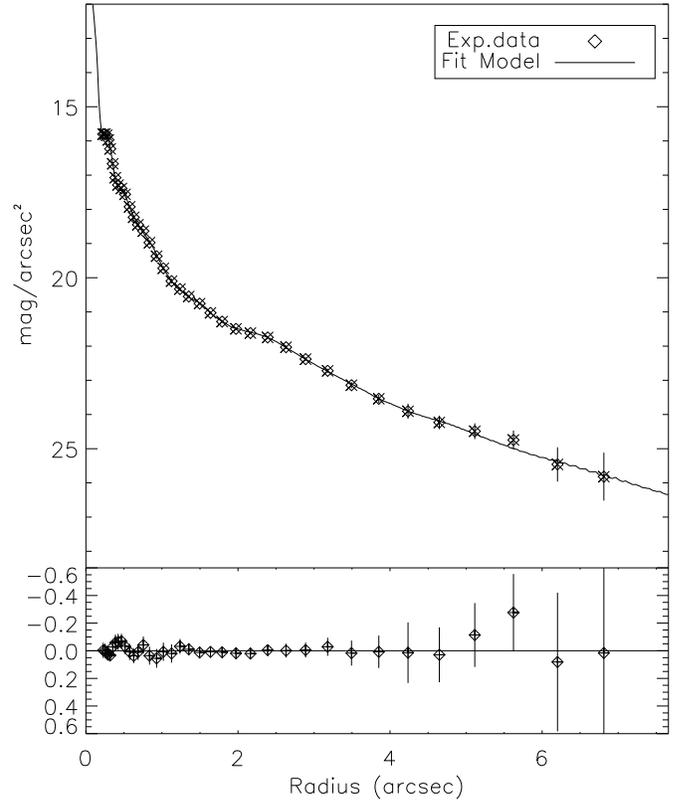}}
\caption{The radial surface brightness profile of PG 1553+11 as
derived from the HST + WFPC2 image (F702W filter). The observed
profile (filled dotted) is compared with the scaled PSF profile.
\label{RadProf} }
\end{figure}

In this framework, the $\chi^{2}$ is no longer a maximum--likelihood
estimator and the statistics cannot provide a way to compute the
$\chi^{2}$ value corresponding to an upper limit of a given
significance level, but $\chi^{2}$ minimization is still applicable
to best--fitting. Since our aim is to evaluate a limit on the
host--galaxy magnitude, we need to take the contribution of
systematic errors into account.

The best fit of the data with the PSF (Fig.3) yields
$\chi^{2}$=1.127, and the rms of the residuals is 0.06 mag. We
consider this quantity as the 1$\sigma$ global uncertainty of the
fit. Applying this 1$\sigma$ variation to the magnitude of the
scaled PSF (0.06 mag brighter), we derived a fit with
$\chi^{2}$=1.141. If a 3$\sigma$ variation is considered, the
$\chi^{2}$ of the fit is 1.216.

We considered all fits to the data with $\chi^{2}<$1.216
(corresponding to the variation at the 3$\sigma$ level) equivalent
to the fit with a simple PSF (object unresolved). In order to
evaluate the upper limit to the brightness of a possible host
galaxy, we computed the $\chi^{2}$ of a two--component fit of the
data (PSF+host galaxy) as a function of the magnitude of the host
galaxy (with effective radius 10 kpc) at given redshifts. An example
in Fig. 4 is shown for $z=0.35$. There is a sharp monotonic increase
of the $\chi^{2}$ for increasing galaxy luminosities . We take the
magnitude corresponding to $\chi^{2}$ at the 3$\sigma$ limit as
upper limit of the host magnitude . In the case reported in the
figure, this corresponds to a maximum magnitude of the host R =
18.07. The procedure was repeated for redshifts in the range from
0.05 to 0.9. This produces a relation between the upper limit of the
absolute host magnitude and the redshift (see Fig.5).

The curve divides the [$M_{R}$,z] plane in two regions: one with
permitted values (host not detectable) above the line and one
forbidden below. The same procedure has been applied assuming a host
galaxy with radius 5 Kpc, with very similar results.

In the transformation of apparent magnitude into absolute one, we
assumed galactic extinction $A_{R}$=0.14; k-corrections derived from
\citet{10} and a cosmology with $H_{0}$=70 Km s$^{-1}$ Mpc$^{-1}$,
$\Omega_{m}= 0.3$ and $\Omega_{\Lambda}= 0.7$.

        \begin{figure}
\centering \resizebox{8cm}{8cm}{\includegraphics{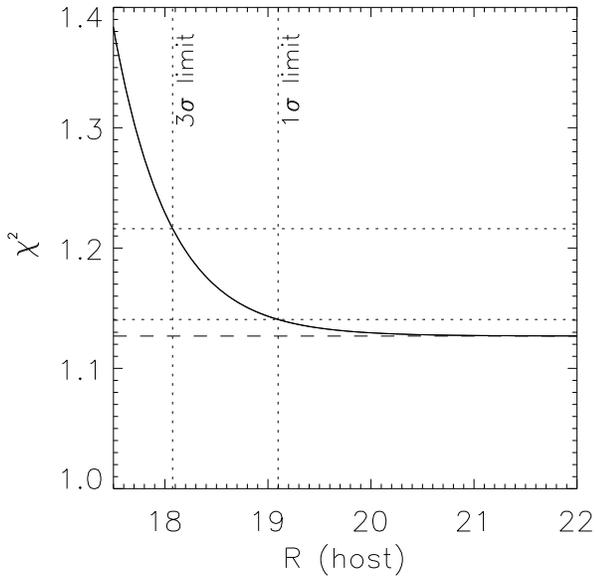}}
\caption{$\chi^{2}$ versus magnitude of the host galaxy of the BL
Lac object PG1553+11 (R=14.4), for a host galaxy with $R_{e}$ = 10
kpc at redshift 0.35.} \label{ChiVar}
\end{figure}

\subsection {Determination of the redshift lower limit}

Based on the limit of the host galaxy magnitude, it is possible to
infer  a lower limit to the distance of the source, because it was
shown \citep{11} that the host galaxy luminosity is encompassed in a
relatively narrow range for all BL Lac objects resolved with HST.
The magnitude distribution is well--fitted by a Gaussian peaked at
$M_R \simeq -22.8$ with FWHM of 1 mag, which is reported in Fig. 5.
If the host galaxy of PG 1553+11 is fairly typical, it is apparent
from the figure that its redshift must be $\geq 0.25$.


As mentioned above, the HST images of PG 1553+11 derive from a
systematic study of 110 BL Lacs \citep{14,13}. In 14 cases, one
being PG1553+11, the redshift is unknown and no indication of a host
galaxy was found. For all the objects, a limit to the magnitude of
the host galaxy was obtained  supposing that the dominant errors
were statistical. Therefore a $3\sigma$ limit could be deduced by
searching for a $\chi^{2}$ variation $\Delta\chi^{2} = 6.6$ where
two variable parameters are considered, i.e. the apparent magnitude
of the host galaxy and its radius.  In the case of PG 1553+11, a
limit $m_R > 21.6$ was obtained by considering a de Vaucouleurs
galaxy with a 10 kpc effective radius. Correspondingly a redshift
limit z$>$0.79 was proposed by \citet{11}.

Compared with our estimate (see Fig.4), the difference in the limit
magnitude of the host galaxy is $\sim$ 2.5 mag,  which is ascribed
to the assumption that the uncertainties were dominated by
statistical errors.

\begin{figure}
\centering \resizebox{8cm}{8cm}{\includegraphics{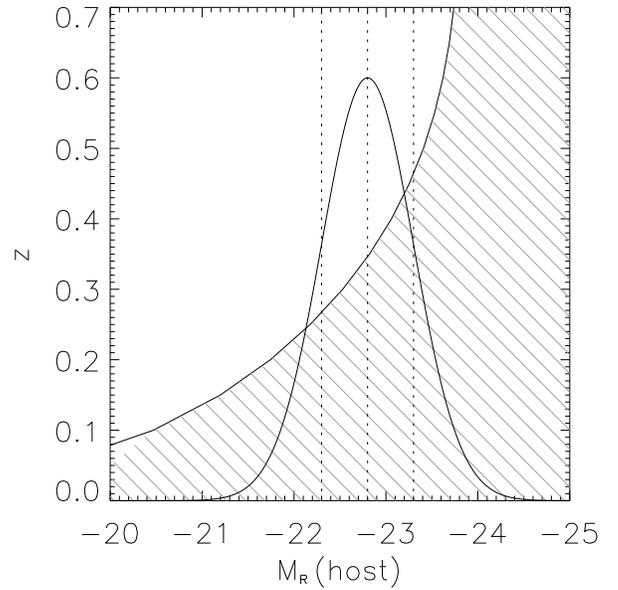} }
\caption{Relation between the upper limit of the host galaxy
($R_{e}$=10 kpc) magnitude and the redshift for the BL Lac object PG
1553+11. Permitted parameters are in the area above the curve. The
Gaussian envelope of the BL Lac host magnitude distribution is also
shown for reference \citep{11}. } \label{ChiVar}
\end{figure}

\section{Discussion}

Comparing our lower limits on the redshift with those deduced from
TeV data, we are led to propose a redshift z= 0.3-- 0.4, making PG
1553+11 the most distant TeV source detected thus far.  The
absorption of EBL would therefore be severe \citep{07}.

From the optical point of view, PG 1553+11 belongs to a restricted
group of very bright objects ($R<16$), for which neither the
redshift nor the host galaxy is known. Similar cases are
\object{0048--099}, \object{1722+119} and \object{2136--428}
\citep[e.g.][]{12}. For all of them, not detecting of the host
galaxy places stringent limits on the redshift. They are all obvious
targets for a search in the TeV band. Their detection would be a
major result and, at the same time, may yield an upper limit to the
distance.

\begin{acknowledgements}

This work was partially supported by the Italian Ministry for
University and Research (MIUR) under COFIN 2002/27145, and ASI--INAF
I/023/05/0. We are grateful to Massimo Persic for discussion of the
TeV observations.

\end{acknowledgements}

\bibliographystyle{aa}

\end{document}